# Efficient fabrication of high-density ensembles of color centers via ion implantation on a hot diamond substrate

E. Nieto Hernandez[1,3#], G. Andrini[2,3,5#], A. Crnjac[4], M. Brajkovic[4], F. Picariello[2,5], E. Corte[1,3,5], V. Pugliese[1,3], M. Matijević[4,1], P. Aprà[1,3], V. Varzi[1], J. Forneris[1,3,5], M. Genovese[,5,3], Z. Siketic[4], M. Jaksic[4], S. Ditalia Tchernij[1,3,5]*

[1] Physics Dept., University of Torino, Italy.

[2] Dipartimento di Elettronica e Telecomunicazioni, Politecnico di Torino, Torino, Italy.

[3] Istituto Nazionale di Fisica Nucleare sez. Torino, Italy.

[4] Laboratory for Ion Beam Interactions, Ruđer Bosǩović Institute, Zagreb, Croatia.

[5] Division of Quantum Metrology and Nanotechnologies, Istituto Nazionale di Ricerca Metrologica (INRiM), Torino, Italy.

[#]: these authors contributed equally to this work.

**Abstract**

Nitrogen-Vacancy (NV) centers in diamond are promising systems for quantum technologies, including quantum metrology and sensing. A promising strategy for the achievement of high sensitivity to external fields relies on the exploitation of large ensembles of NV centers, whose fabrication by ion implantation is upper limited by the amount of radiation damage introduced in the diamond lattice. In this works we demonstrate an approach to increase the density of NV centers upon the high-fluence implantation of MeV $N^{2+}$ ions on a hot target substrate (>550 °C). Our results show that, with respect to room-temperature implantation, the high-temperature process increases the vacancy density threshold required for the irreversible conversion of diamond to a graphitic phase, thus enabling to achieve higher density ensembles. Furthermore, the formation efficiency of color centers was investigated on diamond substrates implanted at varying temperatures with MeV $N^{2+}$ and $Mg^+$ ions revealing that the formation efficiency of both NV centers and magnesium-vacancy (MgV) centers increases with the implantation temperature.

## 1. Introduction

Nitrogen-vacancy (NV) centers have been widely studied in the last decade due to their unique opto-physical properties [1]–[3]. The coupling of their electronic spin to external interacting fields, the optical readout of their spin state based on photoluminescence intensity detection, and the availability of dynamic control protocols enable optically-detected magnetic resonance (ODMR) experiments and make this color center appealing for applications in the fields of quantum computing [4], [5], sensing [6]–[10] and metrology [11], [12]. Several sensing schemes have been proposed, relying on high-density ensembles of NV centers to enable the development of field sensors with unprecedented sensitivity [13], [14] thanks to the maximization of the signal to noise ratio. Additional classes of diamond color centers have also received increasing interest for field sensing because of their narrow zero-phonon line (ZPL) and higher photon emission rate and spin addressability [15]–[19]. Among them, the magnesium-vacancy (MgV) center [16], [19] has recently been demonstrated to exhibit intriguing properties such as a high emission rate and a short excited state lifetime.

Moreover, its field-sensitive spin state has been theoretically predicted to be addressable for operation as a quantum bit for computing and sensing applications [20].

To date, ion implantation followed by high-temperature annealing is the most straightforward method to achieve the localized fabrication of color centers in diamond. NV centers can be routinely fabricated upon the introduction of nitrogen ions in the crystal lattice with a high degree of control on the volume density from the single-ion level [21], [22] to highly dense ensembles of emitters [23]. However, an upper limit to the density of NV centers achievable by ion implantation is set by the concurrent introduction of radiation-induced lattice damage in the diamond crystal [24], [25]. Indeed, the cumulation of radiation-related defects upon keV and MeV ion implantation results in the progressive amorphization of the crystal, whose structure is irreversibly converted to a graphitic phase when the lattice vacancies overcome a threshold volume density (*graphitization threshold* in the following) of ~$1\times10^{22}$ and ~$9\times10^{22}$ ions cm$^{-3}$ for ions in the keV [25] and MeV [26] energy range, respectively. While this feature has been extensively exploited for the diamond lithography processes based on controlled graphitization at the micrometer scale for device and material engineering [27]–[29] it may alter the emission properties of ion-implanted environment-sensitive color centers due to the stray effects related to strain, local charge densities, or spin-spin interactions [30].

Furthermore, the formation efficiency of implanted ions into optically-active color centers is commonly in the 1-10% range for NV centers [31], unless rather articulated processes steps are adopted, such as co-implantation of selected dopants in order to tune the charge environment [32][33] or electron irradiation with in-situ thermal annealing [34], [35].

Both the graphitization threshold and the sub-optimal conversion efficiency concur to define an upper limit to the density of color centers manufacturable in diamond upon conventional ion implantation at room temperature. In this paper we investigate the effects of MeV nitrogen implantation in a hot diamond substrate (500 °C and 750° C temperatures) with respect to the results achievable at room-temperature. We show that the proposed hot substrate implantation process can increase the value of the graphitization threshold, thus enabling the achievement of higher implantation fluences (and thus of higher density NV centers ensembles) without introducing irreversible amorphization in the crystal structure of diamond.

## 2. Experimental

*Samples fabrication*

The experiments were performed on two different samples for NV and MgV centers fabrication, respectively. Sample #1 consisted of a $3\times 3 \times 0.3$ mm$^3$ IIa CVD single crystal diamond produced by ElementSix, which was classified as "optical grade" due to its nominal concentrations of nitrogen and boron of < 1 ppm and < 0.05 ppm, respectively. Sample #2 was a 2 x 2 x 0.5 mm$^3$ "electronic grade" quality substrate, characterized by < 5 ppb substitutional N and B concentrations.
Prior to ion implantation both samples were masked with a thick (> 150 µm) metal layer, in which several apertures of $100 \times 200$ µm were milled to be exploited as implantation masks. Fig. 1 reports a schematic representation of the experimental configuration adopted. This choice enabled the ion implantation processes to be performed on well-defined regions of the same sample under different conditions of temperature, thus enabling their homogeneous assessment and comparison.

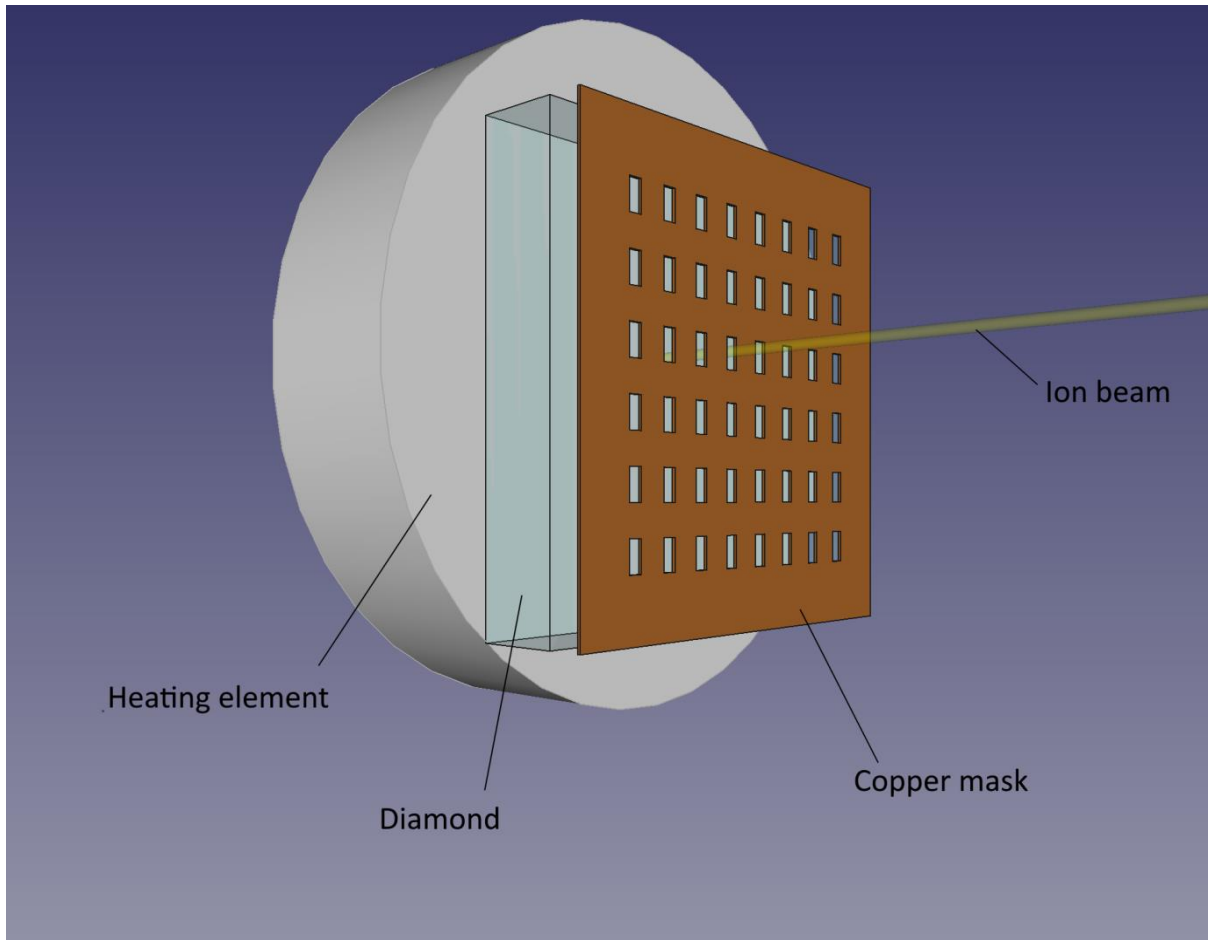

**Figure 1** schematic representation of the adopted configuration for ion implantation on a hot diamond substrate. (not in scale)

Sample #1 was implanted with 2 MeV $N^{2+}$ ions at 4 different fluences: $1\times10^{15}$ cm$^{-2}$, $5\times10^{15}$ cm$^{-2}$, $1\times10^{16}$ cm$^{-2}$, and $2.5\times10^{16}$ cm$^{-2}$. Three regions were implanted for each of the aforementioned implantation fluences, differing for the temperature at which the diamond substrate was held. The temperatures adopted for the ion implantation were the following: (~ 22 °C), 550 °C and 750 °C. Moreover, the time scale at which the ion implantation was performed was defined in order to avoid the introduction of the radiation damage in the crystal lattice at a too high rate (i.e., by adopting a high ion current), thus limiting the diffusion and recombination of the vacancies with respect to the duration of the implantation process. All of the ion implantations were therefore performed on a timescale of 1 h by adjusting the ion current accordingly. In order to suitably satisfy this criterion, the range of the adopted ion beam currents was of $5\times10^{-11}$ - $15\times10^{-11}$ A. The implantations were performed using an ion micro-beam having ~10 µm diameter, which was raster scanned over the diamond surface through the mask apertures by means of electromagnetic focusing and scanning in high vacuum conditions (~$10^{-6}$ mbar). Technical details about the typical irradiation procedure, fluence calibration and sample heating system can be found in [36].

After the ion implantation process, a photoluminescence (PL) characterization of the sample "as implanted" (i.e., prior to any subsequent thermal annealing process) was performed by means of confocal Raman spectroscopy using a Horiba-Jobin-Yvon LabRam HR-VIS instrument equipped with a Peltier-cooled CCD detector array (532 nm laser excitation; 21.6 mW laser power; 20× air objective [37]). Subsequently, the sample underwent an additional annealing step to promote the activation of NV centers from regions implanted at both room- and high-temperature. The annealing was performed at 950 °C for a total processing time of 12 hours. The duration was chosen to be one order of magnitude higher than the one used during the fabrication process.

Sample #2 was implanted with a single fluence of 2 MeV Mg ions ($10^{14}$ cm$^{-2}$), well below the graphitization threshold. The implantation has been performed at 4 different temperatures, in order to assess the effect of the in situ annealing as well as the lowest temperature at which these emitters are optically activated. To this purpose the sample was firstly implanted at 706 °C and then the implantation has been repeated at the same fluence at 502 °C, 300 °C and room temperature (~ 22 °C).

## 3. Results

### 3.1. *NV centers and graphitization*

By comparing the vacancies densities obtained using the Monte-Carlo simulation software SRIM [24] with the graphitization threshold of diamond for MeV ions implantation [26], the four implantation fluences adopted for the fabrication of Sample #1 were chosen to be in proximity of the graphitization threshold (Fig. 2a). The fluences of $1\times10^{15}$ cm$^{-2}$, $5\times10^{15}$ cm$^{-2}$ resulted in a vacancy density profile not exceeding the graphitization threshold. Conversely, the Bragg's peak of the vacancy density profiles corresponding to the $1 \times 10^{16}$ cm$^{-2}$, and $2.5 \times 10^{16}$ cm$^{-2}$ ion fluences exhibits values above the graphitization threshold at the end of the ion range (1 µm from the sample surface). In this latter case, the irreversible formation of a graphitic layer upon a thermal annealing performed above 800 °C is expected if the ion implantation occurs at room temperature conditions [26].

In Fig 2.b an optical micrograph of the different implanted areas is shown for the "as implanted" Sample #1. As expected, the two regions implanted at the highest implantation fluences at room temperature exhibit a significant optical opacity, indicating the amorphization of the irradiated areas. The area implanted at $5 \times 10^{15}$ cm$^{-2}$ fluence appears heavily damaged as well. Conversely, in the diamond regions implanted at 550 °C and 750 °C temperatures, no amorphization of the crystal lattice is apparent on the basis of the optical opacity of the sample for any of the adopted fluences.

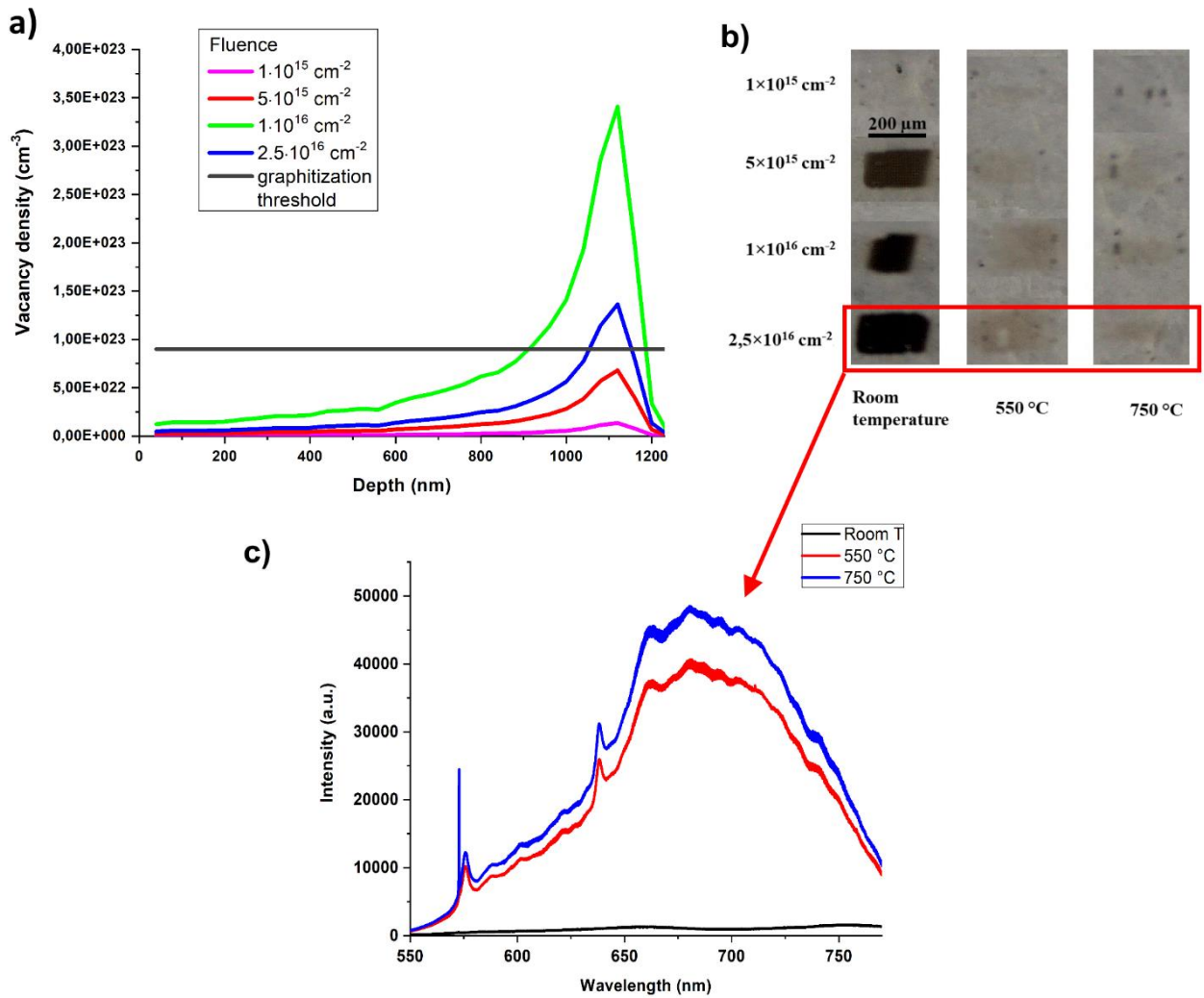

**Figure 2 a)** Vacancy density profiles obtained from SRIM Monte Carlo simulations as a function of the depth for the implanted ion fluences. **b)** Optical micrograph of the areas implanted at different temperatures before ex-situ thermal processing. **c) PL** spectra acquired from the three regions implanted with a fluence of 2,5 × $10^{16}$ cm$^{-2}$ at different temperatures after 12 hours of thermal annealing at 950°C.

### 3.2. *Spectral characterization - NV centers*

In order to assess the formation of optically active NV centers in the implanted regions after the processing of Sample #1, PL spectra were acquired.

Fig. 2c shows a selected set of three PL spectra acquired from the regions of Sample #1 implanted with a fluence of 2,5 × $10^{16}$ cm$^{-2}$ at room temperature (black line); 550°C (red line) and 750°C (blue line) and processed with an additional thermal annealing step of 12 h at 950 °C. Care was taken to preserve the same experimental conditions during the spectral characterization; the same optical excitation power was used (21.6 mW), and the sample was carefully positioned each time, maximizing the PL signal for a consistent PL analysis upon focusing on the sample surface. All the spectra exhibit only three features, namely: the first-order Raman scattering at 572.6 nm (1332 cm$^{-1}$ shift under 532 nm laser excitation), and the ZPLs of the neutral (575 nm) and negative (638 nm) charge states of the NV centers with their own phonon replicas at higher wavelengths, respectively.

A quantitative comparison among the set of the processed regions was performed by integrating the whole PL spectrum related to NV emission, (i.e. in the 560-775 nm range, as highlighted by the yellow area in Fig. 3a). Based on the assumption of identical experimental conditions for each investigated region, the integral value of each PL spectrum is therefore regarded as proportional to the NV concentration in the probed sample volume (~ 2 µm$^3$ under the adopted optical configuration). This assumption is further justified by the lack of spectral features different from the NV emission in the considered PL spectra. A comprehensive set of the acquired PL spectra integral values is shown in Figure 3 against the two relevant experimental parameters (i.e. the implantation temperature and the ion fluence) for the "as implanted" sample (Fig. 3b) and after the 12 h 950 °C post-implantation annealing (Fig. 3c).

Concerning the "as implanted" sample spectral characterization, in the adopted fluence range, Fig. 3b outlines a more intense NV emission from the regions implanted at the lowest considered ion fluence ($1 \times 10^{15}$ cm$^{-2}$). This observation is consistent with what was reported in [31], [38], [39] and imputed to a decrease in the quantum efficiency for higher density of NV centers, related to a "self-absorption" process. Additionally, higher temperature implantations generally yield a higher NV PL intensity with respect to their room temperature counterparts.

The lack of emission in the PL characterization for the regions implanted at room temperature with implantation fluences of $1\times10^{16}$ cm$^{-2}$ and $2.5\times10^{16}$ cm$^{-2}$ after 12h of subsequent annealing (Fig. 3c), is interpreted as a proof of the conversion of the amorphized diamond crystal structure into graphite upon thermal processing. This result is fully in line with what typically reported for implantation fluences above the graphitization threshold [40]. Conversely, the implantation at 550 °C and 750 °C resulted in a *in-situ* recovery of the crystal lattice, thus significantly increasing the graphitization threshold. The increase of this threshold is testified by the unequivocal observation of the NV center spectral feature even at the highest considered implantation fluence.

To summarize, the investigation in the total PL intensity acquired from the sample highlighted two different aspects: firstly, an increase graphitization threshold above the $3.5 \times 10^{23}$ cm$^{-2}$ vacancy density value at both 550 °C and 750 °C implantation temperatures; secondly, the progressively increasing PL intensity at increasing implantation temperatures for the same implantation fluences. This latter observation is indicative of an increased creation efficiency of NV centers upon N implantation. By assuming a linear dependence in the total measured PL intensity on the number of optically active centers, we estimated an increased formation efficiency up to 26%. This estimation was made comparing the PL intensity acquired from the area implanted with the lowest ion fluence ($1 \times 10^{15}$ cm$^{-2}$) at room temperature with that fabricated at 750 °C. This result consistent with what reported in [41].

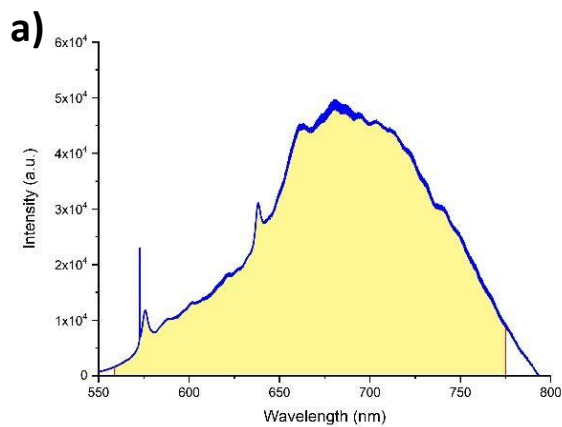
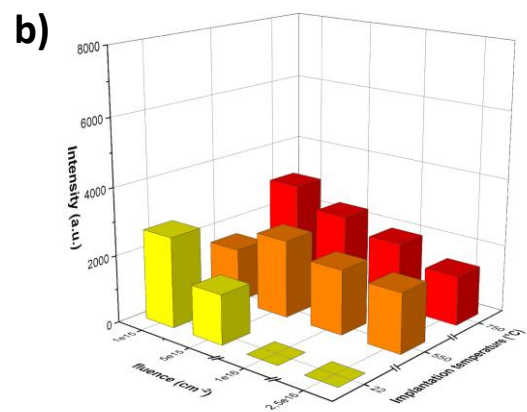

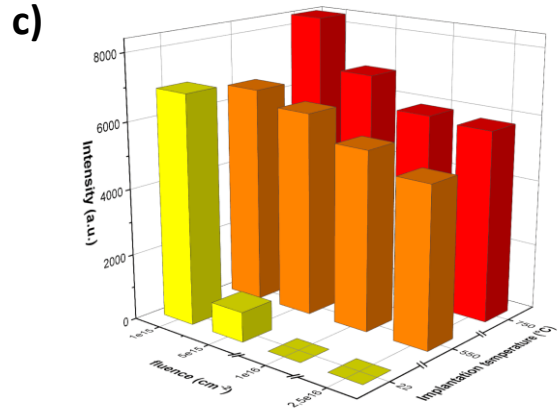

**Figure 3 a) PL** spectrum acquired from the region implanted at room temperature with a fluence of $1\cdot10^{16}$ ions·cm$^{-2}$. The yellow region, delimited by the red lines represent the integrated area. **b,c)** histograms of the intensities obtained from the different implanted region at different subsequent annealing steps. Respectively: **b)** pre annealing °C **c)** 12 h at 950 °C. The intensity scale is maintained fixed in all of the histograms for an easier comparison.

### 3.3. Spectral characterization - MgV centers

Following the Mg$^+$ ion implantation, the emission properties of the "as implanted" Sample #2 were first investigated to address the radiation-induced effects and potential Mg-related spectral features. A preliminary characterization was thus performed under 520 nm CW laser excitation (200 µW) by means of a custom fiber-coupled single-photon sensitive confocal microscope (100x air objective, 0.9 NA) [19], where a long-pass dichroic mirror and a set of optical filters allowed the detection of the fluorescence above 550 nm wavelengths. The spectral analysis was specifically carried out using a single grating monochromator (1200 grooves·mm$^{-1}$, 600 nm blaze, ~ 4 nm spectral resolution) out-coupled via multimode fiber to a single photon avalanche detector (SPAD). The PL spectra reported in in Fig. 4 shows that only the region implanted at 706°C reveals the typical MgV spectral features, denoted by a sharp ZPL at 557 nm and its phonon replicas in the 560-650 nm range. The hot implantation performed at 502°C leads additionally to a weak emission at 557 nm and a stronger photoluminescence signal at 741 nm, respectively denoted as the MgV ZPL and the GR1 band related to the neutral vacancy emission [42]. The appearance of this spectral feature is justified by the fact that the GR1 center has maximum formation yield in this temperature range. Conversely, the PL spectra acquired from the regions implanted at 300 °C and at room temperature exhibit the presence of the GR1 emission only, without any noticeable contribution related to the MgV center. This evidence is in agreement with the commonly adopted experimental protocols for fabricating several diamond color centers, whose conversion into stable optically active defects requires the recombination of the implanted impurities with the lattice vacancies introduced upon ion implantation. This process is activated by supplying sufficiently high thermal energy to achieve vacancy diffusion and results in the disappearance of the GR1 spectral feature at temperatures above 600 ° [42].

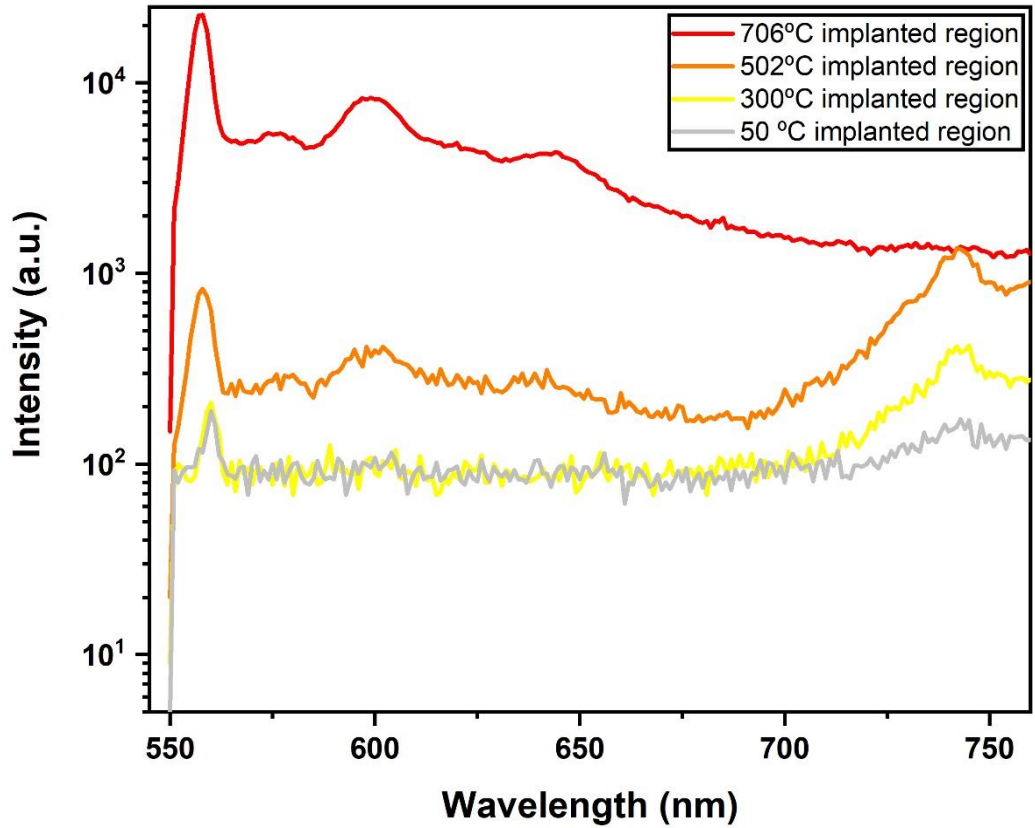

**Figure 4** PL spectra acquired under 520 nm laser excitation (200 µW) from Sample #2 following Mg ion implantation. The four spectra refer to different implantation temperatures: 50 °C (black line), 300 °C (yellow line), 502 °C (orange line) and 706 °C (red line).

After the "as implanted" characterization, the sample was thermally processed to monitor the formation efficiency of MgV centers as a function of the annealing duration. For this reason, the entire process was carried out in multiple steps. The sample was heated up to 900 °C four times with a process duration of 2 h, 4 h, 3 h and 3 h respectively, resulting in a total cumulative annealing time of 12 h.

PL spectra were acquired using the same Raman spectroscopy setup as in the NV analysis to correlate the formation efficiency of the MgV center. The PL spectra acquired from each region after 12 h ex-situ annealing at 900 °C is shown in Fig. 5a. All the considered spectra exhibited the sole MgV spectral features without any contribution from the GR1 center.

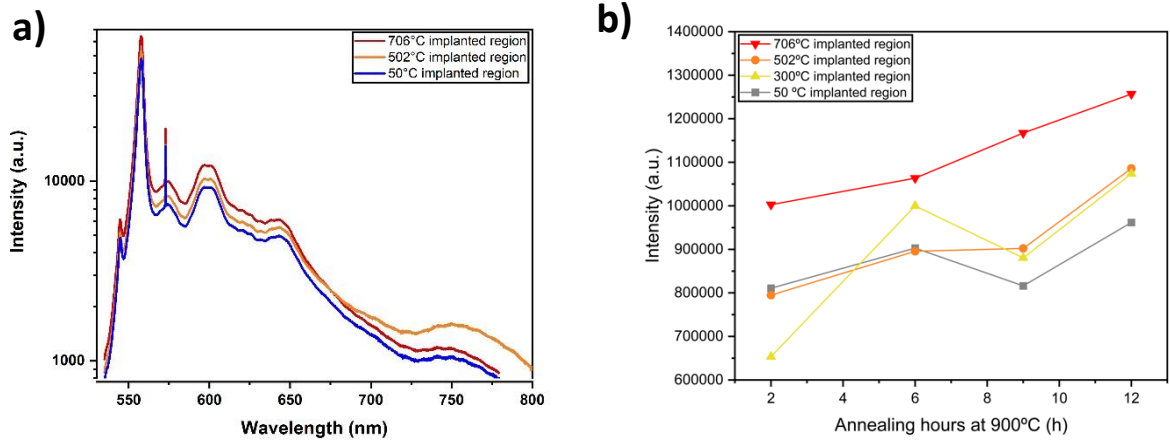

**Figure 5. a)** PL spectrum acquired from the Mg+ implanted region with a fluence of $10^{14}$ cm$^{-2}$ at increasing temperatures, additionally annealed ex-situ at 900 °C for 12 h **b)** 2D representation of the intensities collected from each of the regions after each of the 900°C annealing treatments, each point represent the integral of the spectra in the 530-700 nm range.

Fig. 5b reports the integrated intensities in the 530-700 nm range as a function of the number of annealing hours at the fixed annealing temperature of 900 °C for each implantation temperature. It is noticeable that the overall intensity for the sample area implanted at 706 °C (red triangles) exhibits the highest PL intensity for all the annealing duration steps. In particular, comparing the regions implanted at 502°C and 706°C, an increase in photoluminescence of ~20% is observed for the latter one after the 12 h thermal annealing.

Conversely, the regions implanted at lower temperatures do not exhibit a significant dependence on the implantation temperature itself; each of the 50°C, 300 °C and 502 °C implanted regions show comparable emission intensities at each considered post-implantation annealing step (Fig. 5b). Consequently, this observation suggests that all implantation temperatures are below the threshold required for an enhancement of the MgV center formation efficiency. This indicates that the ion fluence is the sole relevant processing parameter at implantation temperatures below 500 °C.

These results show the advantage of ion implantation on a hot substrate rather than the same process at room temperature to achieve better defect formation rates especially at higher. Similarly long-established procedures for dopants activation in other classes of semiconductor material [43], [44] show that hot implantation leads to better electronic properties and a higher crystal quality overall. Moreover, the reported benefits of the technique open a way to optimize the fabrication of highly dense ensembles of color centers and contribute to further studies on hot implantation processes applied to the optical defects engineering in semiconductors and their optical activation.

Conclusions

In this work we demonstrated the possibility of increasing the graphitization threshold upon MeV nitrogen implantation at high temperature. The upper limit for vacancy density introduction in ion implanted diamond was increased to values larger than $9 \times 10^{23}$ cm$^{-3}$ by performing ion irradiation at sample temperatures above 500 °C. The significant increase in the number of ions that can be introduced in the diamond before irreversibly graphitizing the substrate suggests the possibility of creating larger ensembles of color centers that could increase the system's sensitivity. Additionally, we reported a study on the formation efficiency of both the NV center and MgV center, showing, in both cases, an increment in the photoluminescence of the regions implanted at increasing temperatures with respect to what observed from regions implanted at room temperature. By assuming a linear dependance of the PL intensity with the number of formed optically active centers, it was possible to estimate an increased formation efficiency of ~26% (~20% ) respectively for the NV (MgV) center when implanted at a temperature of 750 °C (706 °C) with respect to conventional room-temperature processes. These values are also expected to depend on other processing parameters, including the implantation fluence and the ion beam current. Also, the same trend was observed both for an impurity-vacancy (NV center) and a split-vacancy (MgV center) defect configuration. This observation suggests that the observed behavior can be expected in general for the formation of extrinsic color centers in diamond. However, the formation enhancement associated with high-temperature ion implantation might depend on the specific ion species and defect structure. Further works will be required to achieve a precise assessment of the role of such parameters in the efficient manufacturing of diamond-based color centers dependance is far beyond the scope of the present work. The reported results are of critical importance for creating large ensembles of color centers in diamond, which can lead to unprecedented field sensitivity for color centers based sensing protocols in future experiments while not sacrificing the spatial resolution by increasing the sensing volume.

*Acknowledgements*

This work was supported by the following projects: 'Intelligent fabrication of QUANTum devices in DIAmond by Laser and Ion Irradiation' (QuantDia) project funded by the Italian Ministry for Instruction, University and Research within the 'FISR 2019' program; 'Training on LASer fabrication and ION implantation of DEFects as quantum emitters' (LasIonDef) project funded by the European Research Council under the 'Marie Skłodowska-Curie Innovative Training Networks' program; experiments ROUGE and PICS4ME, funded by the 5th National Commission of the Italian National Institute for Nuclear Physics (INFN); Project "Piemonte Quantum Enabling Technologies" (PiQuET), funded by the Piemonte Region within the "Infra-P" scheme (POR-FESR 2014-2020 program of the European Union); "Departments of Excellence" (L. 232/2016), funded by the Italian Ministry of Education, University and Research (MIUR); "Ex post funding of research - 2021" of the University of Torino funded by the "Compagnia di San Paolo".
The projects 20IND05 (QADeT), 20FUN05 (SEQUME) and 20FUN02 (PoLight) leading to this publication have received funding from the EMPIR programme co-financed by the Participating States and from the European Union's Horizon 2020 research and innovation programme. The EU RADIATE Project (proposal 19001744-ST) for granting transnational access to the RBI laboratories for Ion Beam interactions.